# Oral messages improve visual search

Suzanne Kieffer
Henri Poincaré University & LORIA
LORIA, BP 239
F54506 Vandoeuvre-lès-Nancy Cedex
Suzanne.Kieffer@loria.fr

Noëlle Carbonell
Henri Poincaré University & LORIA
LORIA, BP 239
F5506 Vandoeuvre-lès-Nancy Cedex
Noelle.Carbonell@loria.fr

## ABSTRACT
Input multimodality combining speech and hand gestures has motivated numerous usability studies. Contrastingly, issues relating to the design and ergonomic evaluation of multimodal output messages combining speech with visual modalities have not yet been addressed extensively.

The experimental study presented here addresses one of these issues. Its aim is to assess the actual efficiency and usability of oral system messages including some brief spatial information for helping users to locate objects on crowded displays rapidly and without effort.

Target presentation mode, scene spatial structure and task difficulty were chosen as independent variables. Two conditions were defined: the visual target presentation mode (VP condition) and the multimodal target presentation mode (MP condition). Each participant carried out two blocks of visual search tasks (120 tasks per block, and one block per condition). Scene target presentation mode, scene structure and task difficulty were found to be significant factors. Multimodal target presentation mode proved to be more efficient than visual target presentation. In addition, participants expressed very positive judgments on multimodal target presentations which were preferred to visual presentations by a majority of participants. Besides, the contribution of spatial messages to visual search speed and accuracy was influenced by scene spatial structure and task difficulty. First, messages improved search efficiency to a lesser extent for 2D array layouts than for some other symmetrical layouts, although the use of 2D arrays for displaying pictures is currently prevailing. Second, message usefulness increased with task difficulty. Most of these results are statistically significant.

## Categories and Subject Descriptors
H.5.2 [**User Interfaces**]: Ergonomics, Evaluation/Methodology, Graphical User Interfaces, Natural Language, Voice I/O.
I.3.6 [**Methodology and Techniques**]: Interaction Techniques.

## Keywords
Visual search. Multimodal system messages. Speech and graphics. Usability study. Experimental evaluation. Visual target spotting.

## 1. CONTEXT AND MOTIVATION
Input multimodality combining speech and hand gestures has motivated numerous usability studies. Contrastingly, to our knowledge, issues relating to the design and ergonomic evaluation of multimodal output messages combining speech with visual modalities, mainly 2D or 3D graphics, have not yet been addressed extensively. Until recently, main research efforts have been focused on the implementation of speech either as a substitute for text in the design of multimedia documents, or as a useful alternative (or supplementary) output medium for both blind (or ill-sighted) users and mobile users of PDAs, wearable computers or embedded systems.

Speech and graphics appear as useful output modalities. First, speech is the most natural human communication modality. Second, most current interactive applications use graphics as their main output modality. Recent scientific advances in the area of conversational user interfaces [3] are liable to stimulate research aimed at endowing interactive systems with human-like multimodal communication capabilities. In particular, numerous prototypes of human-like embodied conversational agents (ECAs) have been developed, ranging from talking heads to real robots.

The main aim of the work presented here is to assess the actual efficiency and usability of speech as a supplementary output modality to standard visual presentations. We performed an experimental study with a view to determining whether oral messages including coarse information on the locations of graphical objects on the current display may facilitate visual search tasks sufficiently for making it worth while to integrate such messages in graphical user interfaces. In addition, we varied display spatial layout in order to test the influence of visual presentation structure on the contribution of these messages to facilitating visual search on crowded displays.

Objectives, working hypotheses, methodology and experimental setup are described in the two following sections. Results are then presented and discussed. Conclusions and future work direction are summed up in the last section.

## 2. OBJECTIVES
Results of an earlier experimental study [1] indicate that coarse spatial information presented orally facilitates visual search for visually familiar components of realistic scenes, compared to situations where the target is visually familiar but its location in the scene is unknown. In addition, they suggest the possible influence of the scene spatial structure on the effectiveness of spatial information messages.

The experiment reported here contributes to validating these preliminary results. It is focused on the search for "(visually) familiar" targets, that is, pre-viewed targets (or "items"). Its main objectives are to ascertain that:

(i) providing coarse information on target location facilitates visual search for familiar targets on crowded graphical displays significantly;
(ii) the resulting improvement in search efficiency (i.e., target selection time and accuracy) is sufficient for motivating designers to integrate oral messages including such spatial information into graphical user interfaces;
(iii) information on target location is useful whatever the layout (or spatial structure) of the scene components, and whatever the difficulty of the search task;
(iv) multimodal system messages associating speech with graphics will meet with general user acceptance.

These objectives are grounded on two working hypotheses inferred from general knowledge on visual perception:

A. By narrowing visual search space, oral messages including information on target location (TL) will sensibly reduce target selection time and improve target spotting accuracy. If the size of the reduced search space is inferior or equal to the size of the human visual field, selection time may be drastically reduced without loss of accuracy, due to the possible occurrence of target "pop out" effects [7].
B. Assuming that visual search strategies are influenced by display layout, the effects of scene spatial structure on users' scan paths may interfere with those of TL information; therefore, search efficiency might vary with scene spatial structure.

## 3. METHOD

The overall experimental protocol is first presented. Then, the description focuses on the design of the visual material.

Target presentation mode, scene spatial structure and task difficulty were chosen as independent variables. Two dependent variables were used to assess participants' performances: target selection time (from scene display onset until first mouse click), and accuracy (i.e., mouse click on the target vs elsewhere); in addition, participants' subjective judgements were elicited through post-session questionnaires and debriefing interviews. To assess hypothesis A, two conditions were defined: the VP condition (target visual presentation) and the MP condition (target multimodal presentation). Each participant carried out two blocks including 120 visual search tasks each, one block per condition. In addition, the order of blocks was counterbalanced between participants so as to neutralize possible task learning effects. To assess hypothesis B, we had to create specific visual material, due to the great structural diversity and complexity of real objects and scenes. To control spatial structure variations, scenes were build from sets of photographs, each scene including 30 photographs arranged along one out of four standard symmetrical structures (see figure 1):

- Matrix-like, the layout most frequently used for presenting collections of pictures [7];
- Elliptic (two concentric ellipses), a useful structure for displaying several sets of pictures simultaneously;
- Radial (8 radii along the screen medians and diagonals), another possible structure for visualizing sets of pictures, especially ordered sets; \item
- Random (random placing of items), the reference layout.

We used the same 120 scenes for the VP and MP conditions in order to eliminate target visual and semiotic characteristics as possible factors of influence on participants' performances. We used 3600 different photographs collected from popular sites on the Internet to build the required 120 scenes (30 photographs per scene), so as to enhance the realism and attraction of the experimental visual search tasks as well as to obtain useful results for the design of picture browsers. Photographs were sorted out into 40 themes (e.g., sport, monuments, animals) and sub-themes (e.g., snakes or cats, for animals), then formatted (125x95 pixels, i.e., 4/3). Scenes were exclusively composed of photographs belonging to the same theme or sub-theme so as to reduce intra-scene diversity in visual salience and subjective appeal. They were presented to participants in random order, regardless of their structure, their visual properties and those of the target. In addition, a different order was assigned to each subject so as to neutralize possible sequence effects. Target position varied from one scene to the other.

**Figure 1: Matrix, Elliptic, Radial and Random structures.**

For each scene, participants had to locate a pre-viewed photograph in the scene, and to select it as fast as they could, using the mouse. In the VP condition, the isolated target was displayed in the centre of the screen during 3 seconds. In the MP condition, a short oral message containing information on the TL was played simultaneously with the target visual presentation. Messages were composed of one or two short spatial phrases, for instance, "On the left (of the screen)" or "At the bottom (of the screen), on the right". Following target presentation, participants had to click on a button in the centre of the screen for launching the scene display. Thus, the position of the mouse at the beginning of the search was identical for all tasks. Three levels of task difficulty were defined, based mainly on the target visual complexity and the number of photographs in the scene that might be mistaken for it because of their visual similarity to it [2]. Levels of task difficulty were evenly distributed among the four structures (i.e., 10 scenes by level and structure).

A gender-balanced group of 24 experienced computer users with ages between 24 and 29 and normal eyesight (assessed using the Bioptor test kit) participated in the experiment. Thus, all participants were expert mouse users with alike quick motor reactions; they were also experienced in visual search activities on computer displays. Therefore, target selection time and spotting accuracy were likely to reflect visual search performance reliably, and task learning effects were prevented.

# 4. RESULTS

Averaged selection times, error numbers and percentages were computed over all subjects (24) by condition, scene structure and task difficulty. First, we applied a n-factors ANOVA procedure on the data, then, paired t-tests whenever possible.

## 4.1 ANOVA Procedure

Table 1 shows that scene target presentation mode, scene structure and task difficulty are significant factors. Considering selection times, results are highly significant for both target presentation mode and task difficulty; considering error numbers, they are highly significant for the target presentation mode only. These results suggest that scene structure has less influence on results than target presentation mode and task difficulty.

**Table 1. ANOVA Procedure.**
Factors: target presentation mode, scene structure, task difficulty.

| Factors | Selection times | Error numbers |
|---|---|---|
| Presentation | t=1202.98; p<.0001 | t=23.18; p<.0001 |
| Structure | t=6.26; p=0.0003 | t=2.58; p=0.0005 |
| Difficulty | t=32.49; p<.0001 | t=7.59; p=0.0005 |

## 4.2 Multimodal vs Visual Target Presentation

Spatial information messages improved participants' visual search performances significantl2. Actually, averaged target selection times computed over all participants are thrice longer in the VP condition than in the MP condition (5674 ms versus 1747 ms). This result is highly significant (t=-34.07; p<.0001). Selection times and error numbers per condition are reported in tables 2 and 3. Average selection times (Avg ST) and standard deviations (Std Dev) were computed over the total number of tasks per condition (Nb Obs).

Moreover, participants expressed very positive judgments on multimodal target presentations, both in the questionnaires and during the debriefing interviews. For 75\% of them (18), target spotting had been easier (less hesitations) in the MP condition than in the VP reference condition. Most participants mentioned that they had experienced some strain and visual fatigue during the VP condition whereas they had felt perfectly comfortable during the MP condition. All participants considered that oral messages including coarse information on target location could provide efficient support to visual search activities, and two thirds (16) expressed a marked preference for the MP condition.

**Table 2. Participants' selection times per condition.**

| Condition | Avg ST ms | Std Dev ms | Nb Obs |
|---|---|---|---|
| VP | 5674 | 5985 | 2880 |
| MP | 1747 | 1552 | 2880 |

**Table 3. Participants' errors per condition.**

| Condition | Nb Errors | % Errors | Nb Obs |
|---|---|---|---|
| VP | 150 | 5.2% | 2880 |
| MP | 79 | 2.7ù | 2880 |

These quantitative and qualitative results confirm those presented in [4, 5] for more complex tasks and other combinations of modalities: speech+graphics versus text+graphics. They partly validate hypothesis A: additional oral information on the location of a visually familiar target on the display significantly improves visual search efficiency effectively. Such messages also increase visual search comfort, and will get wide user acceptance.

## 4.3 Effects of Scene Structure

In the reference condition (VP), the four spatial structures can be ordered as follows according to increasing averaged selection times: Radial (5626 ms), Random, Matrix, Elliptic (6250 ms). Selection time differences between the Radial and Elliptic structures, the Radial and Matrix structures, the Elliptic and Matrix structures are statistically significant; see table 4 where values (720 tasks per condition and structure) preceded by "-" or "+" are respectively inferior or superior to the corresponding average values per condition reported in table 1. These results are somewhat unexpected, since participants were experienced computer users, and the use of 2D arrays is currently prevailing for displaying pictures. For the MP condition, the ranking of the four structures is the same as for the VP condition (see table 4) but only the difference between the Radial and Elliptic structures reaches statistical significance (t=2.75; p=.006).

**Table 4. Participants' selection times per condition and structure.**

| Structure | Condition | Avg ST ms | Std Dev ms |
|---|---|---|---|
| Radials | VP | -5081 | -5565 |
|  | MP | -1640 | -1256 |
| Random | VP | -5626 | -5819 |
|  | MP | -1737 | -1437 |
| Matrix | VP | +5738 | -5879 |
|  | MP | +1763 | +1819 |
| Elliptic | VP | +6250 | +6585 |
|  | MP | +1851 | +1633 |

**Table 5. Participants' errors per condition and structure.**

| Structure | Condition | Nb Errors | % Errors |
|---|---|---|---|
| Radials | VP | 34 | 24% |
|  | MP | 9 | 14% |
| Random | VP | 29 | 21% |
|  | MP | 17 | 26% |
| Matrix | VP | 36 | 25% |
|  | MP | 16 | 24% |
| Elliptic | VP | 42 | 30% |
|  | MP | 24 | 36% |

Concerning accuracy (see table 5), "actual errors" only are considered in this subsection and the next one. They include mouse clicks on non targets and clicks on the background (i.e., targets not found); clicks near the target (22) are considered as hits. In the VP condition, rates of actual errors range from 21\% (Random structure) to 30\% (Matrix structure) of the total number of actual errors (141). Differences between structures are then moderate. Contrastingly, there is a sharp difference between the Radial structure (14\% over a total of 66 actual errors) and the three other structures (from 24\% to 36\%) in the MP condition. A likely interpretation of this unexpected result is that the zones defined on the screen by the chosen spatial phrases match the Radial structure best and the Elliptic one worst. This interpretation is supported by some spontaneous comments collected during the debriefing interviews. Participants' subjective judgments are at variance with their performances. In the VP condition, more than half of the participants expressed a marked preference for Elliptic layouts compared to the other structures, and two thirds of them judged either the Matrix or the Radial structure the most inefficient layout. In the MP condition, judgments were more varied: the Radial and Elliptic structures were preferred by 11 and 8 participants respectively, while the Matrix and Elliptic structures were viewed as most inefficient by 7 and 6 participants respectively. Participants' performances and subjective judgments concerning the Matrix structure in the VP condition are in accordance with the results presented in [6].

These quantitative and qualitative results suggest two main conclusions. First, messages including information on target location facilitate visual search for familiar pictures or graphical objects whatever the display layout. However their efficiency may be reduced in cases when spatial phrases and scene spatial structure imply different partitions of the display. This result contributes to validating hypothesis B. Therefore, display layout should be taken into account when designing verbal messages meant to help users to spot familiar pictures or graphical objects on crowded displays. Second, participants' performances and judgments relative to the VP condition suggest that 2D arrays may prove to be less appropriate than some other symmetrical layouts for displaying small collections of pictures or graphical objects. Further experimental research is needed to ascertain this conclusion which, if proved to be valid, might induce designers of graphical user interfaces and picture browsers to question the current prevailing use of 2D arrays for designing display layouts.

## 4.4 Effects of Task Difficulty

Participants' selection times validate our classification of scenes into three levels of difficulty (40 scenes per condition and level). In both conditions, averaged selection times increased noticeably from level 1 (Easy) to level 3 (Very Difficult). For the VP condition, the difference between any pair of levels is statistically significant, the difference between levels 1 and 3 being highly significant ($t=-6.40$; $p<.0001$). For the MP condition, differences between level 1 and 3, and 2 and 3, are highly significant ($t=-5.29$; $p<.0001$ and $t=-5.33$; $p<.0001$ respectively), while the difference between levels 1 and 2 did not reach significance. Error rates also increased from level 1 to 3 in both conditions.

A careful analysis of participants' performances shows that average selection times increase from level 1 to level 3 less rapidly in the MP condition (25%) than in the VP condition (35%). This observation suggests that spatial information messages are particularly useful for performing difficult visual search tasks. These results contribute to validating the second part of hypothesis A. Therefore, it seems worth while to assist users in difficult visual search activities through spatial information messages. As such short oral messages will be well accepted by potential users, or so it seems according to participants' subjective judgments, their use for helping users to carry out easy visual search tasks may also be considered.

## 5. CONCLUSION AND FUTURE WORK

We have presented an experimental study that aims at assessing the actual contribution of voice system messages to visual search efficiency and comfort. Messages comprised one or two spatial phrases conveying coarse information on the target location on the display. 24 participants carried out 240 visual search tasks in two conditions differing from each other in initial target presentation only: visual presentation of the target \textit{versus} multimodal presentation, that is, visual presentation of the target simultaneously with oral indications on its location on the screen. Oral messages improved participants' selection times and accuracy noticeably. However, their influence varied according to display spatial layout: the benefits were smaller for 2D array layouts than for Radial layouts, although the use of 2D arrays for displaying pictures is currently prevailing. In addition, message usefulness increased with task difficulty. Most of these results are statistically significant. According to subjective judgments, oral messages were well accepted, and multimodal target presentations were preferred to visual presentations by a majority of participants. Therefore, designers of graphical user interfaces might consider resorting to short oral messages including coarse spatial information for drawing users' attention to some displayed object. As such messages are likely to be well accepted by users, they may provide designers of advanced conversational user interfaces with a useful alternative "pointing" technique which may appropriately replace visual enhancement in interaction contexts where gaze activity is intense and where there is a risk of visual attention overload and eyestrain. However, these results need to be consolidated and further refined before reliable recommendations inferred from them can be proposed to designers. Their actual scope has first to be assessed. In particular, is the influence of spatial layout on visual search efficiency and comfort independent of the number of items displayed simultaneously and of their type (e.g., textual labels, graphical icons, drawings, etc.)? To what extent are the efficiency and user acceptance of oral spatial information messages dependent on their length and complexity? We are considering addressing some of these issues in the near future.